\documentstyle[12pt]{article}
\begin{document}

\begin{center}

{\Large \bf Weakly nonlinear quantum transport: an exactly solvable model}

\bigskip

Jian Wang and Qingrong Zheng

\bigskip

{\it
Department of Physics, \\
The University of Hong Kong,\\
Pokfulam Road, Hong Kong.
}

\bigskip

Hong Guo

\bigskip

{\it
Centre for the Physics of Materials,\\
Department of Physics, McGill University,\\
Montreal, Quebec, Canada H3A 2T8.
}

\end{center}

\vfill

\baselineskip 15pt               

We have studied the weakly non-linear quantum transport properties of 
a two-dimensional quantum wire which can be solved exactly. 
The non-linear transport coefficients have been calculated and 
interesting physical properties revealed. In particular we found 
that as the incoming electron energy approaches a resonant point 
given by energy $E=E_r$, where the transport is characterized 
by a complete reflection, the second order non-linear conductance 
changes its sign. This has interesting implications to the current-voltage
characteristics.  We have also investigated the establishment of 
the gauge invariance condition. We found that for systems with a finite 
scattering region, correction terms to the theoretical formalism
are needed to preserve the gauge invariance. These corrections were 
derived analytically for this model.

\vfill

\baselineskip 16pt


\newpage
\section{Introduction}

Non-linear quantum transport in mesoscopic systems has been a very active
research field in recent years\cite{but1,altshu,wingreen,tab}. 
Taboryski {\it et. al.}\cite{tab} have reported observations of non-linear
and asymmetric conductance oscillations of quantum point contacts at a 
small bias voltage. They found that the non-Ohmic and asymmetric
behavior causes a rectified dc signal as the response to an applied ac
current. On the theoretical side, Wingreen {\it et. al.}\cite{wingreen} have
presented a general formulation to deal with the situation of a non-linear
and time-dependent current going through a small interacting region 
where electron energies can be changed by time-dependent voltages. At the 
same time, B\"uttiker and his co-workers\cite{but4,but1,but2} have advanced 
a current conserving theory for the frequency dependent transport. Recently,
this current conserving formalism has been applied to a two dimensional
mesoscopic conductor\cite{wang1}. This 
theory can also be applied to discuss the non-linear behavior of mesoscopic 
samples and the theory is gauge invariant. It has been recognized\cite{but3}
that in non-linear coherent quantum transport, it is essential to consider the 
internal self-consistent potential in order to satisfy the gauge invariant 
condition.  This condition demands that all physical properties predicted by 
a theory can not change if there is a global voltage shift. Obviously this is
a fundamental requirement. 

Recently, Christen and B\"uttiker\cite{but3} have investigated 
the rectification coefficient of a quantum point contact and the
non-linear current-voltage characteristic of a resonant level in a
double barrier structure using the theory of gauge invariant non-linear 
conductance. Another important application of this theory is to 
investigate two-dimensional (2D) mesoscopic and ballistic quantum devices 
which can now be routinely fabricated in many laboratories. Unfortunately 
due to a particular technical difficulty, namely the evaluation of a 
quantity called {\it sensitivity} (see below), so far little is known
for the non-linear conductance in 2D. Clearly an understanding of 2D
situation is very much needed in order to gain further intuition to the 
coherent transport and to predict the non-linear characteristics of the 
variety of 2D nanostructures. The purpose of this paper is to investigate the 
gauge invariant non-linear transport in a specific two dimensional system 
which can be solved exactly.  Hence we were able to obtain various relevant 
physical quantities.  Although a general study for an arbitrary 2D system 
seems difficult, our perspective is that an exactly solved model is 
valuable since it clearly and unambiguously reveals the physical properties 
of the non-linear transport coefficients.  

To be specific, we have considered a very simple 2D model which is a 
quasi-1D ballistic conductor\cite{bag} with a $\delta$-potential confined 
inside, as shown in figure (1a). Because quantum scattering in this system 
leads to mode mixing which is the basic feature of a two-dimensional system, 
it provides answers to our 2D problem.  In a previous work\cite{wang2} we
have used this model to study the electric current conservation of the
AC-transport formalism at the linear conductance level, and calculated the
important physical quantities such as the global and local partial density 
of states.  In the following we shall extend our calculation to explicitly
calculate the second order non-linear conductance $G_{111}$ and $G_{112}$. 
Due to the gauge invariant condition (see below), 
we should have $G_{111}+G_{112}=0$. 
It turns out that for systems with a finite scattering volume, as those of 
any numerical calculations, we found that a correction term must be added to 
satisfy this condition. For this system there is a resonant state with energy 
$E_r$ characterized by a complete reflection, {\it i.e.} the reflection 
coefficient $R=1$. Our results showed that the second order non-linear 
conductance $G_{111}$ changes sign near the resonant point $E_r$.
This leads to interesting current-voltage characteristics of this system.

The paper is organized as the following.  In the next section we shall
briefly review the gauge invariant theory for non-linear transport set
out by B\"uttiker\cite{but1}. In section III we will present the solution 
of the 2D scattering problem. Some of the technical details of section
III have been put into the Appendix. Our results are presented in
section IV. The last section serves as a brief summary.

\section{Gauge invariant formalism}

To be complete, we shall first briefly review the gauge invariant formalism 
of Christen and B\"{u}ttiker\cite{but3} and set out our calculation procedure
for the 2D system. For a multi-probe mesoscopic system, the current through 
probe $\alpha$ is given by\cite{but1,but3}
\begin{equation}
I_{\alpha} = \frac{2e}{h} \sum_{\beta} \int dE f(E-E_F-eV_{\beta})
A_{\alpha \beta}(E, \{V_{\gamma}\}) \\,
\label{e1}
\end{equation}
where $f(E)$ is the Fermi distribution function, and  
\begin{equation}
A_{\alpha \beta}(E, \{V_{\gamma}\}) = Tr[{\bf{1}}_{\alpha} \delta_{\alpha 
\beta} - s^{\dagger}_{\alpha \beta}(E, \{V_{\gamma}\}) s_{\alpha
\beta}(E, \{V_{\gamma}\})]
\label{e2}
\end{equation}
are the screened (negative) transmission functions. For the weakly
non-linear transport, Eq.(\ref{e1}) can be expanded with respect to the
voltages $V_{\beta}$, 
\begin{equation}
I_{\alpha} = \sum_{\beta} G_{\alpha \beta} V_{\beta} + \sum_{\beta
\gamma} G_{\alpha \beta \gamma} V_{\beta} V_{\gamma} + ... \\,
\label{e3}
\end{equation}
where
\begin{equation}
G_{\alpha \beta} = \frac{2e^2}{h} \int dE (-\partial_E f) A_{\alpha
\beta} \\
\label{e4}
\end{equation}
is the linear conductance and 
\begin{equation}
G_{\alpha \beta \gamma} = \frac{e^2}{h} \int dE (-\partial_E f)
(\partial_{V_{\gamma}} A_{\alpha \beta} +\partial_{V_{\beta}} A_{\alpha
\gamma} +e \partial_E A_{\alpha \beta} \delta_{\beta \gamma})\\
\label{e5}
\end{equation}
is the second order non-linear conductance. In Eqs.(\ref{e4}) and
(\ref{e5}), the $A_{\alpha \beta}$ are evaluated at $\{V_{\gamma}\}=0$. 
The requirements that the current is conserved and be independent of a 
global voltage shift (gauge invariance) yield\cite{but1,but3,but5}
\[\sum_{\alpha} G_{\alpha \beta} = \sum_{\beta} G_{\alpha \beta} =0\]
and 
\[ \sum_{\alpha} G_{\alpha \beta \gamma} = \sum_{\beta} G_{\alpha
\beta \gamma} = \sum_{\gamma} G_{\alpha \beta \gamma} = 0\ \ .\]
From this equaiton and Eq.(\ref{e5}), the gauge invariance
condition for $A_{\alpha \beta}$ is
\begin{equation}
e\partial_E A_{\alpha \beta} +\sum_{\gamma} \partial_{V_{\gamma}}
A_{\alpha \beta} = 0 \ \ .
\label{e8}
\end{equation}
The derivative $\partial_{V_{\gamma}} A_{\alpha \beta}$ can be expressed 
in terms of functional derivative of $A_{\alpha \beta}$ with respect the 
electric potential $U$ and the characteristic potential $u_{\gamma}$ which
satisfies $\sum_{\gamma} u_{\gamma} = 1$
\begin{equation}
\partial_{V_{\gamma}} A_{\alpha \beta} = \int d^3 {\bf r} \frac{\delta
A_{\alpha \beta}}{\delta U({\bf r})} u_{\gamma}({\bf r}) \ \ .
\label{e9}
\end{equation}

To gain further insight on Eq.(\ref{e8}), let's consider a two probe
system. Eq.(\ref{e3}) can be written as
\[I_1 = G_{11} V_1 +G_{12} V_2 +G_{111} V_1^2 +2 G_{112} V_1 V_2 + G_{122}
V_2^2\]
Obviously, $G_{12}=-G_{11}$ due to the conservation of electric current. From 
Eqs.(\ref{e5}), (\ref{e8}), and (\ref{e9}) and using the fact that $u_1 +u_2 
=1$, we have $G_{111} = -G_{112} = G_{122}$. Therefore the current depends only
on the voltage differences which is the direct consequence of the gauge 
invariant condition Eq.(\ref{e8}).  We obtain,
\begin{equation}
I_1 = G_{11} (V_1 -V_2) + G_{111} (V_1 -V_2)^2\ \ \ .
\label{iv}
\end{equation}

To calculate the transmission functions $A_{\alpha\beta}$ and their functional
derivatives, we need the characteristic potential $u_{\gamma}$.  This, in
turn, needs the solution of the Poisson equation with a nonlocal screening 
term\cite{but1,but3}. To actually carry out this procedure is very complicated. 
However if we can use the Thomas-Fermi approximation, which is more
appropriate for metallic conductors, the characteristic potential is 
simplified and is found to be related to the local partial density of states.
Within Thomas-Fermi screening, we obtain
\begin{equation}
u_{\gamma}({\bf r}) = \frac{dn({\bf r},\gamma)}{dE}/\frac{dn({\bf r})}{dE} \ \ ,
\label{e10}
\end{equation}
where the partial local density of states $dn({\bf r},\gamma)/dE$ is called
the injectivity and is given by\cite{but5}
\begin{equation}
\frac{dn({\bf r},\gamma)}{dE} = \sum_n \frac{|\Psi_{\gamma n}|^2}{h 
v_{\gamma n}} \ \ \ ,
\label{inject}
\end{equation}
where $v_{\gamma n}$ is the channel velocity and $\Psi_{\gamma n}$ is a 
scattering state. Finally the term $dn({\bf r})/dE =\sum_{\alpha} dn({\bf r},
\alpha)/dE$ is the total local density of states. Substituting Eqs.(\ref{e2}) 
and (\ref{e10}) into Eq.(\ref{e9}), we obtain
\begin{equation}
\partial_{V_{\gamma}} A_{\alpha \beta} = - \int d^3 {\bf r} \eta_{\alpha \beta}
\frac{dn({\bf r},\gamma)}{dE}/\frac{dn({\bf r})}{dE} 
\label{e11}
\end{equation}
where 
\begin{equation}
\eta_{\alpha \beta} = 
s^{\dagger}_{\alpha \beta} \frac{\delta s_{\alpha \beta}}{\delta U({\bf r})} 
+ s_{\alpha \beta} \frac{\delta s^{\dagger}_{\alpha \beta}}{\delta U({\bf r})} 
\label{e12}
\end{equation}
is called {\it sensitivity}\cite{gas1}. We are aware of two ways of calculating
the sensitivity\cite{gas1}. The first is to evaluate $\delta s_{\alpha 
\beta}/\delta U$ directly by introducing a $\delta$-function of infinitesimal 
strength $\delta U$ inside the scattering region. Alternatively, one
can calculate it using the retarded Green's function.  For a 2D system, in
general the Green's function can not be obtained explicitly, hence we shall
use the first method by directly compute the sensitivity. After obtaining the
sensitivity, we can then compute $\partial_{V_{\gamma}} A_{\alpha \beta}$ from
Eq. (\ref{e11}), and obtain $G_{\alpha\beta\gamma}$ from Eq. (\ref{e5}).

\section{Model and Analysis}

As mentioned in the introduction, figure (1a) shows the system where a 
$\delta$-potential is confined inside a quasi-1D wire with width $a$. 
We assume, for simplicity of the calculation, that the boundaries of the 
ballistic conductor are hard walls, {\it i.e.} the potential $V=\infty$ at the
walls. Inside the conductor, the potential is zero everywhere except that 
a $\delta$ function potential $V(x,y) = \gamma \delta(x) \delta(y-y_0)$ is 
placed at position ${\bf r}=(0,y_0)$. The scattering region $x_1 < x < x_2$ 
is assumed to be symmetric with $x_2 = -x_1 =L/2$. From now on we set 
$\hbar= 1$ and $m=1/2$ to fix our units.

The transmission and reflection amplitudes have been calculated using a mode 
matching method\cite{schult,bag}. When the incident electron is in the 
first subband, in an earlier work we have explicitly obtained
these amplitudes\cite{wang2}.  The evaluation was straightforward but quite
tedious, we refer interested reader to Ref. \cite{wang2} for details of
this algebra.  Here we only quote the results: for reflection the amplitude is
\begin{equation}
b_n = \frac{-i \Gamma_{n1}}{2k_n \alpha} \\ ,
\label{bn}
\end{equation}
and for transmission it is 
\begin{equation}
c_n = \delta_{n1} +b_n \ \ .
\label{cn}
\end{equation}
Here $\alpha=1+i\sum_n \Gamma_{nn}/(2k_n)$; $\Gamma_{nm} = \gamma
\chi^*(y_0) \chi_m(y_0)$, and $\chi_n(y)$ is the wave function of the
$n$-th subband along y-direction.  $k_n=E-(n\pi/a)^2$ is the longitudinal 
momentum for the $n$-th mode; $i=\sqrt{-1}$.  Note that for electron 
traveling in the first subband, $k_n$ with $n>1$ is purely imaginary. 
For our coordinate system the scattering matrix elements 
$s_{\alpha \beta}$ are given by $s_{11} = b_1 \exp(ik_1 L)$ and 
$s_{12} = c_1 \exp(i k_1 L)$. 

As mentioned in the last section, to calculate the non-linear conductance of
our 2D sample, it is necessary to find the sensitivity $\eta_{\alpha\beta}$.
Hence according to Eq. (\ref{e12}) we must evaluate 
$\delta s_{\alpha \beta}/\delta U(x_1,y_1)$ where the pair $(x_1,y_1)$ is an
arbitrary location in the scattering volume. For a general 2D sample a direct 
calculation of this functional derivative is very difficult if not 
impossible.  Fortunately for our model this can actually be done exactly.
As a first step we shall introduce an additional $\delta$-potential of
infinitesimal strength $\delta U$ at position $(x_1,y_1)$ inside the
scattering volume. Thus our system becomes to that shown in figure (1b). 
Then we shall solve the scattering matrix formally as a functional of 
$\delta U$. Obviously being able to carry out this step is crucial. 
Finally the functional derivative is performed. 
To proceed we again use the mode matching method\cite{schult,bag,wang2}. 
We will assume $x_1<0$ in the following calculation. The calculation for 
$x_1>0$ can be done in a similar fashion. The electron wave functions are 
written as follows. For region I (see figure. (1b)):
\[\Psi_I = \sum_n \chi_n(y) (a_n e^{i k_n x} + b_n e^{-i k_n x}) \ \ , \]
where $a_n$ is the incoming wave amplitude and taken as an input
parameter; $b_n$ is the reflection amplitude. Similarly for region II
\[\Psi_{II} = \sum_n \chi_n(y) (e_n e^{i k_n x} +f_n e^{-i k_n x}) \ \ ,\]
and for region III
\[\Psi_{III} = \sum_n \chi_n(y) c_n e^{i k_n x} \ \ ,\]
where $c_n$ is the transmission amplitude. We shall match the
wavefunctions and their x-derivatives at the positions $x=x_1$ and 
$x=0$. We obtain, at $x=x_1$ 
\begin{equation}
a_n e^{i k_n x_1} +b_n e^{-i k_n x_1} = e_n e^{i k_n x_1} +f_n e^{-i k_n
x_1} \ \ ,
\label{eq1}
\end{equation}
and
\begin{eqnarray}
ik_n (e_n e^{ik_n x_1} -f_n e^{-i k_n x_1}) & - & ik_n (a_n e^{i k_n x_1}
-b_n e^{-i k_n x_1}) \nonumber \\
& = & \sum_m \tilde{\Gamma}_{nm} (e_m e^{i k_m x_1} +f_m
e^{-i k_m x_1}) \nonumber \\
\label{eq2}
\end{eqnarray}
where $\tilde{\Gamma}_{nm} = \delta U \chi_n^*(y_1) \chi_m(y_1)$. 

At $x=0$, the matching gives 
\[e_n +f_n = c_n \ \ ,\]
and
\[ik_n c_n -ik_n (e_n - f_n) = \sum_m \Gamma_{nm} c_m \ \ .\]
To simplify the notation, from now on $(x_1,y_1)$ is replaced by $(x,y)$.
From the last two equations, we solve for $e_n$ and $f_n$
\begin{equation}
2ik_n e_n = -\sum_m P_{nm} c_m \ \ ,
\label{eq5}
\end{equation}
and 
\begin{equation}
2ik_n f_n = \sum_m \Gamma_{nm} c_m \ \ ,
\label{eq6}
\end{equation}
where $P_{nm} = \Gamma_{nm} - 2i k_n \delta_{nm}$. Eliminating $b_n$
from Eqs.(\ref{eq1}) and (\ref{eq2}), we obtain
\[2ik_n e_n e^{ik_n x} = 2ik_n a_n e^{ik_n x} +\sum_m \tilde{\Gamma}_{nm}
(e_m e^{ik_m x} +f_m e^{-ik_m x}) \ \ .\] 
Taking the limit $\delta U \rightarrow 0$, we have
\begin{equation}
2ik_n \frac{\delta e_n}{\delta U} = \sum_m \bar{\Gamma}_{nm} (a_m
e^{ik_m x} + b_m e^{-i k_m x}) \ \ ,
\label{eq7}
\end{equation}
where $\bar{\Gamma}_{nm} = \tilde{\Gamma}_{nm}/\delta U$. To arrive at 
the above result we have used the fact that as $\delta U \rightarrow 0$, 
{\it i.e.} when the extra $\delta$ function vanishes, we must have $e_m=a_m$ 
and $f_m=b_m$. From Eq.(\ref{eq5}), 
\begin{equation}
2ik_n \frac{\delta e_n}{\delta U} = -\sum_m P_{nm} \frac{\delta
c_n}{\delta U} \\ .
\label{eq8}
\end{equation}
From Eqs.(\ref{eq7}) and (\ref{eq8}), we arrive at
\begin{equation}
- \sum_m P_{nm} \frac{\delta c_m}{\delta U} e^{i k_n x} = \sum_m
\bar{\Gamma}_{nm}(a_m e^{ik_m x} +b_m e^{-ik_m x} ) = \chi_n^* \Psi \\.
\label{eq9}
\end{equation}
where $\Psi =\Psi_I$ for $x<0$. From Eq.(\ref{eq9}), we have
\[\frac{\delta c_l}{\delta U} = -\sum_n (P^{-1})_{ln} e^{-i k_n x}
\chi_n^* \Psi \ \ .\]
The matrix $P^{-1}$ has been obtained in Ref. \cite{wang2}, and we quote
\[(P^{-1})_{ln} = \frac{i}{2k_l} (\delta_{ln} -\frac{i \Gamma_{ln}}{2k_n
\alpha}) \ \ .\]
From this equation and Eq.(\ref{cn}) we see that for $l=1$, {\it i.e.}, 
the first subband, $(P^{-1})_{1n} = i c_n/(2k_1)$ provided that $\chi_n$ 
is real which is true in our case. This yields
\begin{equation}
\frac{\delta c_1}{\delta U} = \frac{1}{2ik_1} \sum_n c_n \chi_n
e^{-ik_n x} \Psi \\. 
\label{dc}
\end{equation}

Similarly, from Eqs. (\ref{eq1}), (\ref{eq5}), (\ref{eq6}), (\ref{eq8}), and 
(\ref{eq9}), we obtain
\begin{eqnarray}
\frac{\delta b_n}{\delta U} & = & \frac{\delta e_n}{\delta U}e^{2ik_n x} +
\frac{\delta f_n}{\delta U} \nonumber \\
& = & -\frac{e^{2ik_n x}}{2ik_n} \sum_m P_{nm} \frac{\delta c_m}{\delta
U} + \frac{1}{2ik_n} \sum_m \Gamma_{nm} \frac{\delta c_m}{\delta U} 
\nonumber \\
& = & \frac{\sin(k_n x)}{k_n} \chi_n \Psi + \frac{\delta c_n} 
{\delta U} \nonumber
\end{eqnarray}
When $n=1$, $\delta b_1/\delta U$ becomes, 
\begin{eqnarray}
\frac{\delta b_1}{\delta U} & = & \frac{1}{2ik_1} (\chi_1 e^{ik_1 x} + 
\sum_n b_n \chi_n e^{-ik_n x} ) \Psi \nonumber \\
& = & \frac{1}{2ik_1} \Psi^2 \ \ \ . 
\label{db}
\end{eqnarray}

Because the scattering matrix elements $s_{11}\sim b_1$ and $s_{12}\sim
c_1$ as mentioned above, with the functional derivatives Eqs. (\ref{db})
and (\ref{dc}) we can evaluate $\delta s_{\alpha\beta}/\delta U$ trivially
thus obtaining the sensitivity $\eta_{\alpha\beta}$ of 
Eq. (\ref{e12}). Then using the prescription discussed at the end of
Section 2, we obtain all the weakly non-linear conductances and other
quantities of interest.  Our results will be presented in the next section.

To end this section of the theoretical analysis, we mention that
to check the result of functional derivatives, in the Appendix we shall 
explicitly calculate a quantity called {\it emissivity}\cite{but1} using 
these functional derivatives. In the absence of a magnetic field, it is 
known\cite{but5} that emissivity equals to the injectivity defined in 
Eq.(\ref{inject}) which we can compute using the wavefunctions. Indeed we
confirm in the Appendix that these two equal thus providing a necessary
check to the calculations presented here.

\section{Results}

To obtain numerical results from our analytical formula, for the system of 
figure (1a) we consider incident electron coming from probe 1 and set 
$a=L=1$, $y_0=0.3$, and $\gamma=-1$. Although we have restricted the
incoming electron energy to the first subband, quantum scattering at the
$\delta$-function potential leads to mode mixing.  Thus in our numerical
calculations we have included 50 modes in the scattering volume.  We have
checked that this is enough to obtain good numerical convergence.

As a first result we plot the sensitivity $\eta_{11}({\bf r},E)$ as a 
function of the electron incident energy $E$ at several positions ${\bf r}$. 
This is shown in figure (2).  As discussed in Section 2,
$\eta_{\alpha\beta}$ appears naturally in the theoretical formalism, and it
essentially describes the local electric current response of the scattering
problem when there is a small local potential change. It is related to the 
real part of the diagonal elements of the Green's function\cite{gas1}.  
figure (2) not only shows interesting behavior of this quantity, but also 
gives vivid intuition about the local current response. As shown in our
earlier work\cite{wang2} and mentioned above, the quantum wire studied here
has a resonant state at energy $E=E_r = 36.65$ where we have a complete
reflection (reflection coefficient $R=1$). Not surprisingly, the sensitivity
has a large peak at this resonance energy because the system response is
most sensitive to potential perturbations at resonance. On the other hand,
this peak is larger when we are closer to the $\delta$-function scatterer
located at $x=0$: this indicates that the local perturbation has larger 
effects when it is closer to the scattering center. Although figure (2) shows
$\eta_{11}$ at positions to the left of the scatterer, we have checked that
its behavior is exactly the same for positions $x>0$ as our system is
symmetric.

Adding up all the local responses according to Eq. (\ref{e11}), we can 
explicitly examine the gauge invariant condition Eq.(\ref{e8}). Using
Eqs.(\ref{e2}) and (\ref{e9}) and the fact that $u_1+u_2=1$, 
Eq.(\ref{e8}) reduces to 
\[2 Re(s^{\dagger}_{\alpha \beta} \frac{ds_{\alpha \beta}}{dE}) + 
2 \int d^3 {\bf r} Re(s^{\dagger}_{\alpha \beta} \frac{\delta s_{\alpha
\beta}}{\delta U({\bf r})}) = 0\ \ \ .\]
It is straightforward to evaluate the left hand side of this equation.
Using the functional derivatives obtained in the last section, as well as
the energy derivatives of Eqs. (\ref{bn}) and (\ref{cn}),
we found that the left hand side of the above equation is nonzero, 
and is given by
\begin{equation}
correction = \frac{|s_{12}|^2}{k^2_1} Re(s_{11}) +
Re(\sum_{n=2} \frac{b_1 |b_n|^2} {k_1 k_n} (e^{ik_n L} -2))\ \ .
\label{cor}
\end{equation}
Thus in order to have precise gauge invariance, this correction must be
included. From this result, we notice that the first correction term is only
significant near the first subband threshold where $k_1\approx 0$ and is
negligible for larger incoming electron energies. For the second
correction term, let's examine its behavior near the $n$-th subband with 
$n>1$. From Eq.(\ref{bn}) we see that as the incoming electron 
momentum $k \rightarrow k_n$, $b_1 \rightarrow k_n$ and $b_n$ is finite. 
Therefore, the second correction term remains finite when electron energy 
approaches the $n$-th subband ($n>1$). This is different from the 
AC-transport where the correction diverges\cite{wang2} near the $n$-th 
subband with $n>1$. We emphasize that the correction term is due to the fact 
that we are considering a finite scattering volume. As the scattering volume 
or the incident energy become larger, the effect of these correction terms
diminishes. This can be seen clearly due to the factor $k_1$ in the
denominator, and the exponentially decaying factor $\exp(ik_nL)$ as $k_n$ is
purely imaginary for all $n>1$.  

Now we present numerical evaluations of the analytical formula derived
in the last section for the second order non-linear conductance 
$G_{\alpha \beta \gamma}$. In figure (3) for a comparison the solid line 
shows the linear conductance $G_{21}$ which is proportional to the 
transmission coefficient by the Landauer formula. The dotted line shows 
the second order non-linear conductance $G_{111}$. In the 
presence of an attractive $\delta$-function scatterer, there exists a
quasi-bound state at $E=E_r = 36.65$. As a result, we observe the complete 
reflection\cite{wang3,wang2} indicated by $G_{21}$=0. As expected, the
non-linear coefficient $G_{111}$ also vanishes at $E_r$. Furthermore 
$G_{111}$ changes its sign as the incoming electron energy passes through 
the resonant energy. This has important implications on the current-voltage 
characteristics if we recall the I-V relation Eq. (\ref{iv}). The I-V curves 
of this quantum wire system is shown in figure
(4) for several different electron energies. We can clearly see that when 
$E$ is smaller than $E_r$, {\it e.g.} for $E=10.91$ and $36.22$, since both 
$G_{21}$ and $G_{111}$ are positive, the current $I_1$ increases with 
potential difference of the two probes $\Delta V=V_1-V_2$.
However when $E=37.11$ which is just above $E_r$, $I_1$ is a decreasing
function of $\Delta V$ due to the negative non-linear conductance $G_{111}$.
For this case the quantum wire has a negative differential resistance.
Finally at a even larger energy $E=39.03$, while $G_{111} < 0$, the value
of $G_{21}$ is large enough such that the linear contribution dominates at 
small $\Delta V$ and the non-linear term is larger at larger $\Delta V$.
This behavior is shown in the dash-dotted line in figure (4).

To numerically demonstrate the gauge invariance, we have also computed the 
non-linear conductance $G_{112}$ according to Eq. (\ref{e5}) and it is 
shown in figure (5) as the dotted line. For comparison we have re-plotted 
the non-linear conductance $G_{111}$ (solid line). Note that near the 
resonant region, these two non-linear conductances are very close to each 
other. This is a surprising result, because if the gauge invariant condition 
is precisely satisfied we should have $G_{111}=-G_{112}$. Thus without the 
correction term discussed above, the gauge invariance could not even be 
satisfied up to a sign !  Hence it is important, in any numerical
calculations, to have a large scattering volume\cite{but5}.  
Indeed, when we add the correction term of Eq.(\ref{cor}) to $G_{112}$,
(dashed line in figure (5)), we obtain the expected perfect agreement to the
gauge invariance.

\section{Summary}

To summarize, we have solved exactly the weakly non-linear transport 
characteristics of a two-dimensional quantum wire model. To the best of
our knowledge this is the first exact solution for a truly two-dimensional
ballistic model. The second order non-linear conductances are derived 
analytically. We found that as the incoming electron energy crosses the 
resonant point, the non-linear conductance changes its sign.  This leads to 
interesting current-voltage behavior when the incoming electron energy 
changes. We have also examined the gauge invariant condition which is obtained 
by the global voltage shift. We found that for systems with a 
finite scattering volume, correction terms are needed to preserve the 
gauge invariant condition. We have derived these corrections analytically 
for our model. The correction term consists of two parts. The first part 
dominates when the incident energy $E$ is near the first subband threshold. 
On the other hand the second part is given by the amplitudes of the 
non-propagating modes and is significant near the resonant point. 
Finally, our exact calculation reveals the interesting behavior of the 
sensitivity which describes the local electric current response to a 
potential perturbation.  

\section*{Acknowledgments}

We gratefully acknowledge support by a RGC grant from the Government of 
Hong Kong under grant number HKU 261/95P, a research grant from the 
Croucher Foundation, the Natural Sciences and Engineering Research 
Council of Canada and le Fonds pour la Formation de Chercheurs 
et l'Aide \`a la Recherche de la Province du Qu\'ebec.
We thank the Computer Center of the University of Hong Kong for
computational facilities.

\section*{Appendix}

To check our result of the functional derivatives, {\it i.e.}
Eqs. (\ref{dc}) and (\ref{db}), in this Appendix we compute the 
emissivity defined as\cite{but1}
\[\frac{dn(\alpha, {\bf r})}{dE} = -\frac{1}{4\pi i} \sum_{\beta} Tr [
s^{\dagger}_{\alpha \beta} \frac{\delta s_{\alpha \beta}}{\delta U({\bf
r})} - \frac{\delta s^{\dagger}_{\alpha \beta}}{\delta U({\bf r})} s_{\alpha 
\beta} ] \ \ .  \]
It has been shown\cite{but5} that in the absence of a magnetic field the
emissivity is equal to the injectivity defined in Eq.(\ref{inject}).
We shall explicitly perform the functional derivatives to confirm this
fact, hence provide the necessary check to our algebra.

Using Eqs.(\ref{dc}) and (\ref{db}),  we have
\begin{eqnarray}
s^*_{11} \frac{\delta s_{11}}{\delta U} + s^*_{12} \frac{\delta s_{12}}{\delta 
U} &=& c_1^* \frac{\delta c_1}{\delta U} +b_1^* \frac{\delta b_1}{\delta U} 
\nonumber \\
& = & \frac{1}{2ik_1} (c_1^* \chi_1 e^{-ik_1 x} + c_1^* \sum_n b_n
\chi_n e^{-ik_n x} + b_1^* \chi_1 e^{ik_1 x} + b_1^* \sum_n b_n
\chi_n e^{-ik_n x} ) \Psi \nonumber \\
& = & \frac{1}{2ik_1} (b_1^* \chi_1 e^{ik_1 x} +\chi_1 e^{-ik_1 x} 
+(1+2 b_1^*) \sum_{n=2} b_n \chi_n e^{-ik_n x}) \Psi 
\label{e39}
\end{eqnarray}
where the relation $c_1=1+b_1$ has been used. Before we proceed further,
let us derive a useful relation from the unitary condition of the
scattering matrix, namely
\begin{equation}
1+2b_1^* = -\frac{b_1^*}{b_1} = \frac{\alpha}{\alpha^*} \ \ .
\label{useful1}
\end{equation}
The first equality comes from the unitary condition $c_1^* b_1 + c_1
b_1^* =0$ or $b_1^* + (1+2b_1^*) b_1 = 0$; and the second equality is
from Eq.(\ref{bn}). Since the incoming electron is in the first subband,
we have $k_n^* = -k_n$ for $n>1$. Hence for $n>1$, 

\begin{equation}
\frac{b_n}{b_n^*} = \frac{\alpha^*}{\alpha} \ \ .
\label{useful2}
\end{equation}
Substituting Eqs.(\ref{useful1}) and (\ref{useful2}) into Eq.(\ref{e39}), 
we obtain
\begin{equation}
s^*_{11} \frac{\delta s_{11}}{\delta U} + s^*_{12} \frac{\delta s_{12}}{\delta 
U} = \frac{1}{2ik_1} |\Psi|^2 \ \ ,
\label{e42}
\end{equation}
which is equivalent to Eq. (\ref{inject}). Notice that the imaginary part of 
left hand side of Eq.(\ref{e42}) is proportional to the emissivity. Its real
part gives the sensitivity $\eta_{11} + \eta_{12}$. From the unitary
condition we have $\eta_{11} +\eta_{12}=0$ which agrees with Eq.(\ref{e42}).

\newpage
\section*{Figure Captions}

\begin{itemize}

\item[{Figure 1.}] Schematic plot of the quantum wire system. (a). The 
quantum wire system we have studied: a $\delta$ potential 
$\gamma\delta(\vec{r}-\vec{r_0})$ is confined inside a quasi-1D quantum 
wire, with $\vec{r_0} = (0,y_0)$. The wire width is $a$. The scattering 
region is between $x_1$ and $x_2$, where $x_2=-x_1=L/2$. In our 
calculations, the parameters are set to $L=a=1$, $y_0=0.3$, and 
$\gamma=-1.0$. (b). To compute the functional derivatives of the scattering
matrix with respect to a local potential change, we add another $\delta$
function potential at the position $(x_1,y_1)$.  In this case the system is
divided into three regions by the dotted lines for the boundary matching
solution of the Schr\"odinger equation.

\item[{Figure 2.}] The sensitivity $\eta_{11}({\bf r},E)$ as a function of
energy $E$ at three different positions $x=-L/2, -L/4, 0$ with the same
$y=0.5$. For different $y$ the curve $\eta_{11}$ as a function of $E$ will
be multiplied by a constant. Other system parameters are the same as those of 
figure (1). 
Here the unit of energy is $\hbar^2/(2ma^2)$. 

\item[{Figure 3.}] The conductances $G_{21}$ and $G_{111}$ as functions of
energy $E$.  Solid line: $G_{21}$; dotted line: $G_{111}$.
Other system parameters are the same as those of figure (1).
Here the unit of energy is $\hbar^2/(2ma^2)$. 

\item[{Figure 4.}] The current-voltage characteristics as calculated from
Eq. (\ref{iv}) at several different electron energies $E=10.91, 36.22,
37.11, 39.03$. $\Delta V=V_1-V_2$. Other system parameters are the same as 
those of figure (1).

\item[{Figure 5.}] A numerical check of the gauge invariant condition.
Solid line is $G_{111}$, dotted line is $G_{112}$, dashed line is
the $G_{112} + correction$ where $correction$ is given by Eq. (\ref{cor}).  
Clearly $G_{112}+ correction\ =\ -G_{111}$.
Here the unit of energy is $\hbar^2/(2ma^2)$. 

\end{itemize}
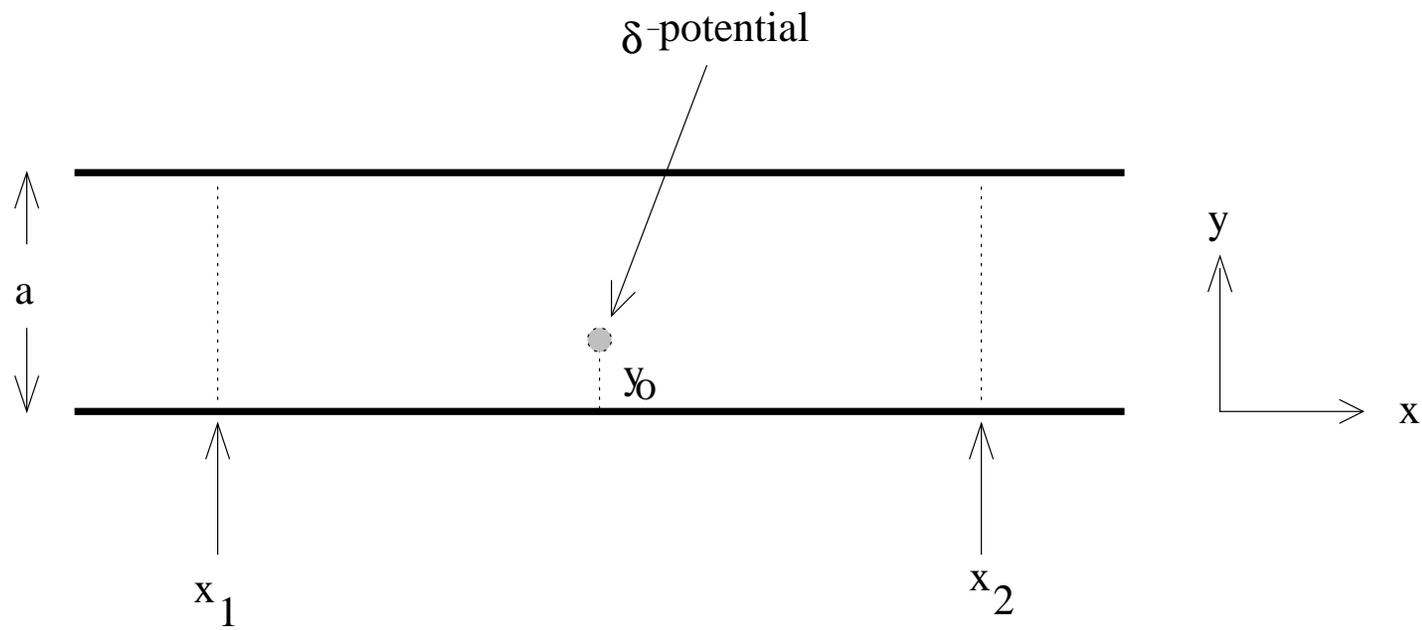
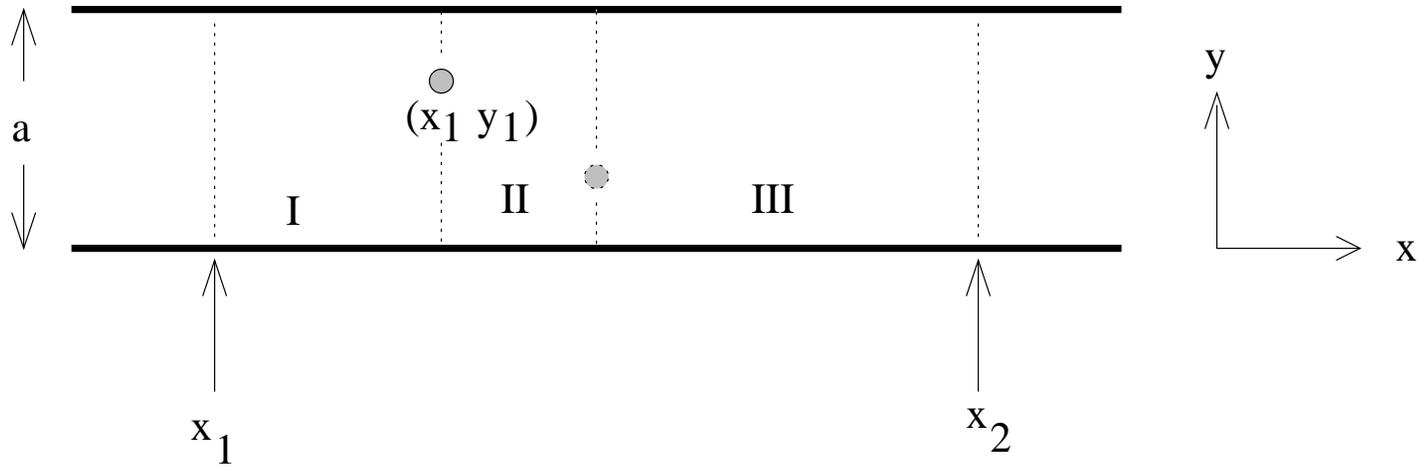
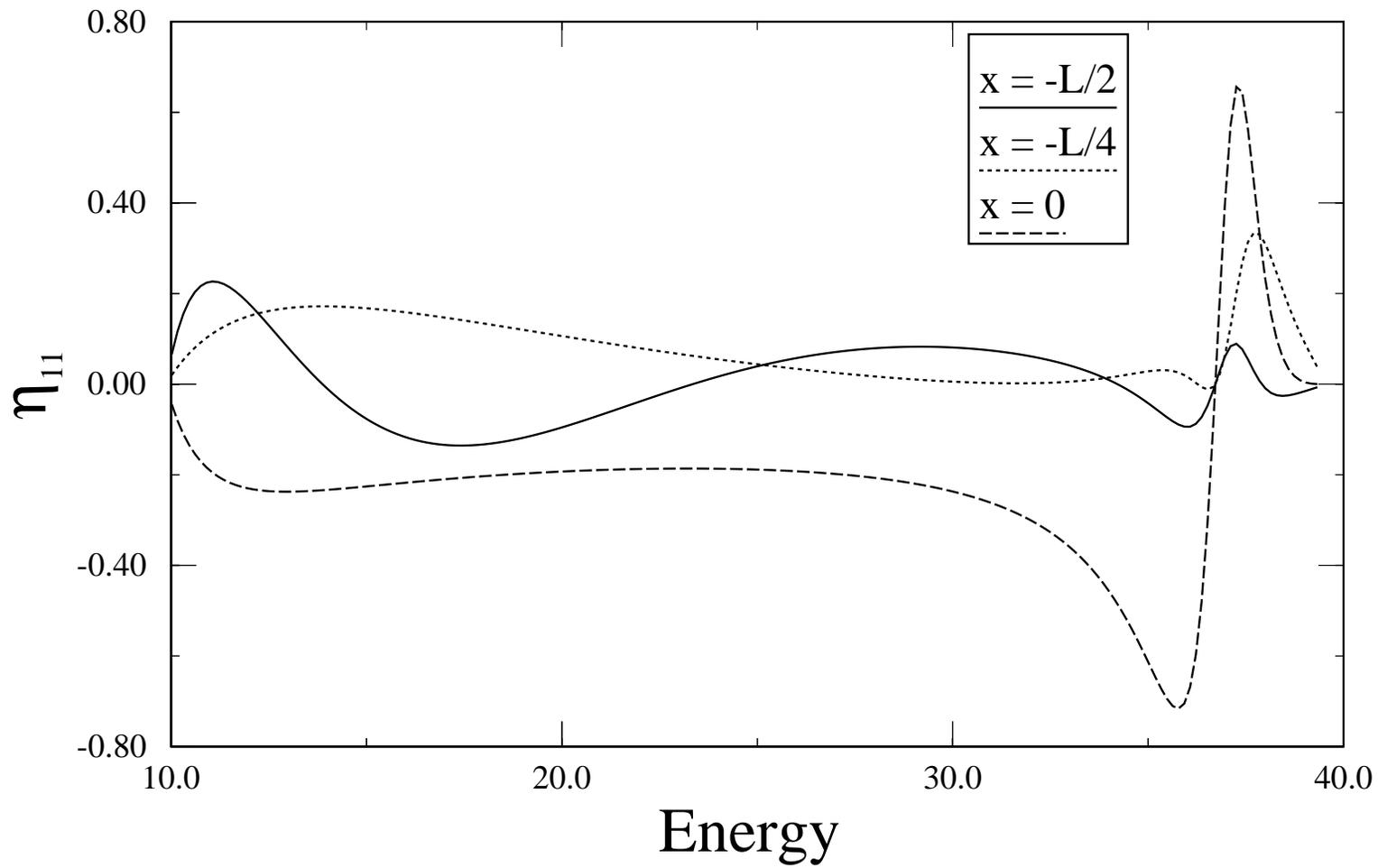
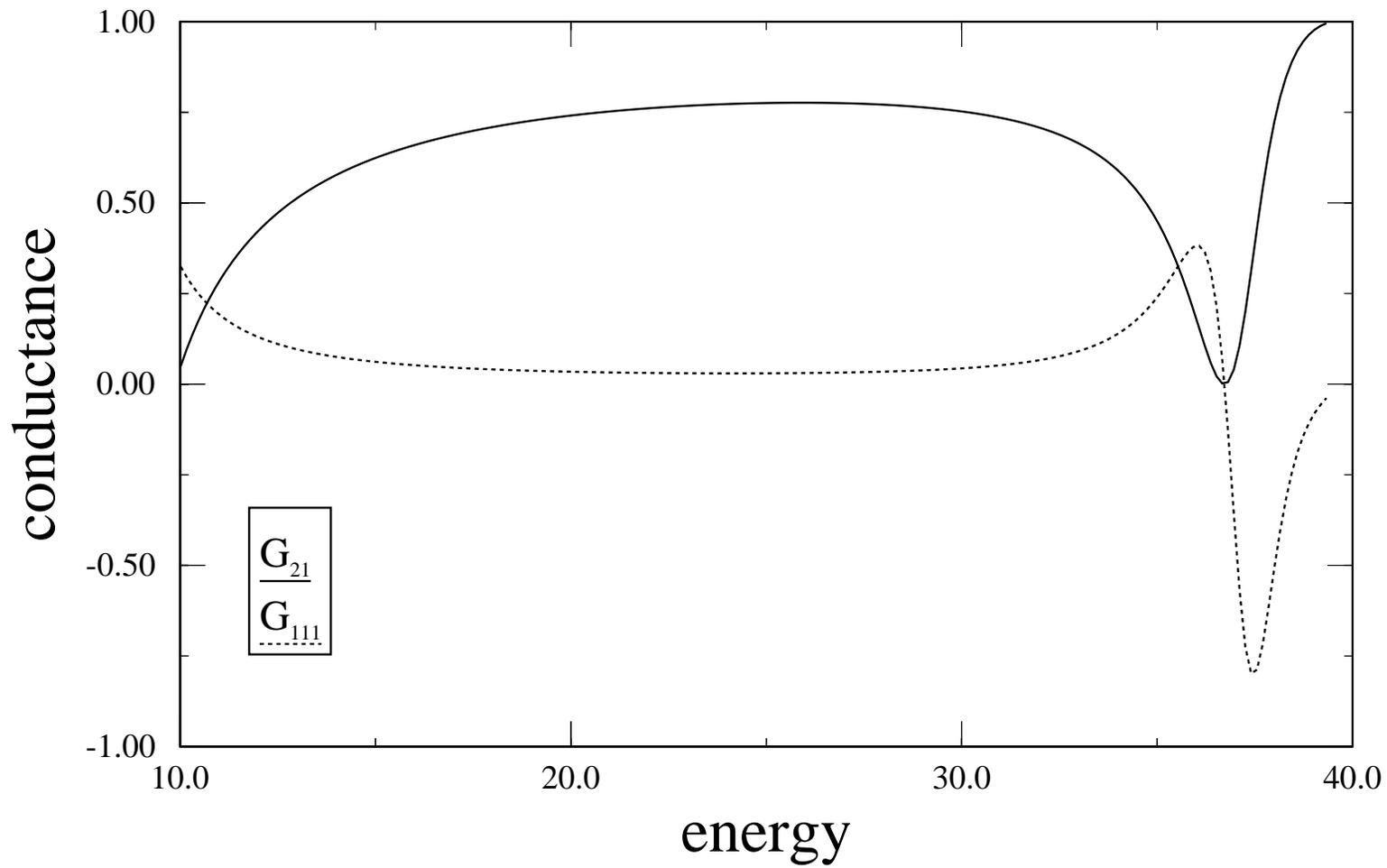
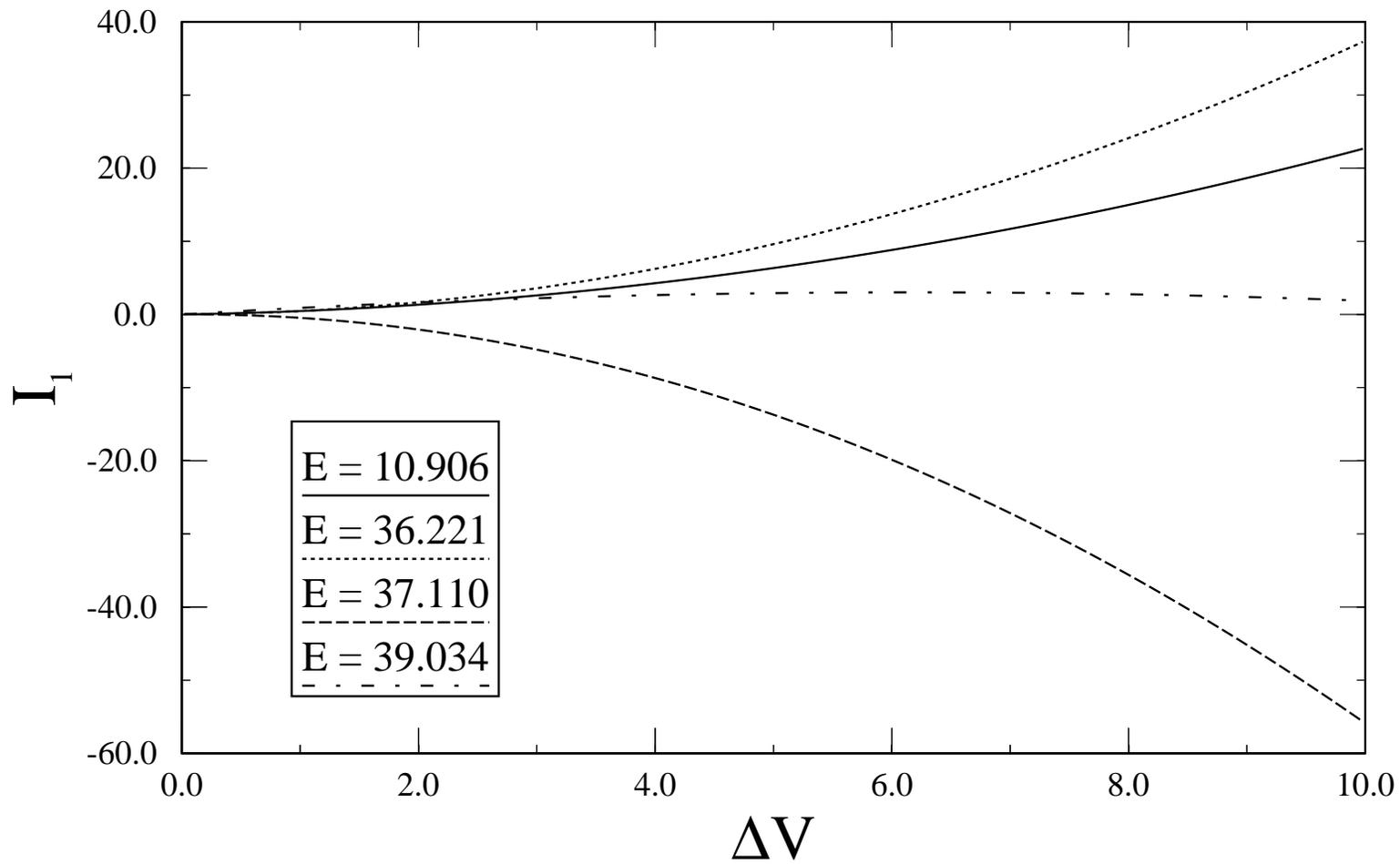
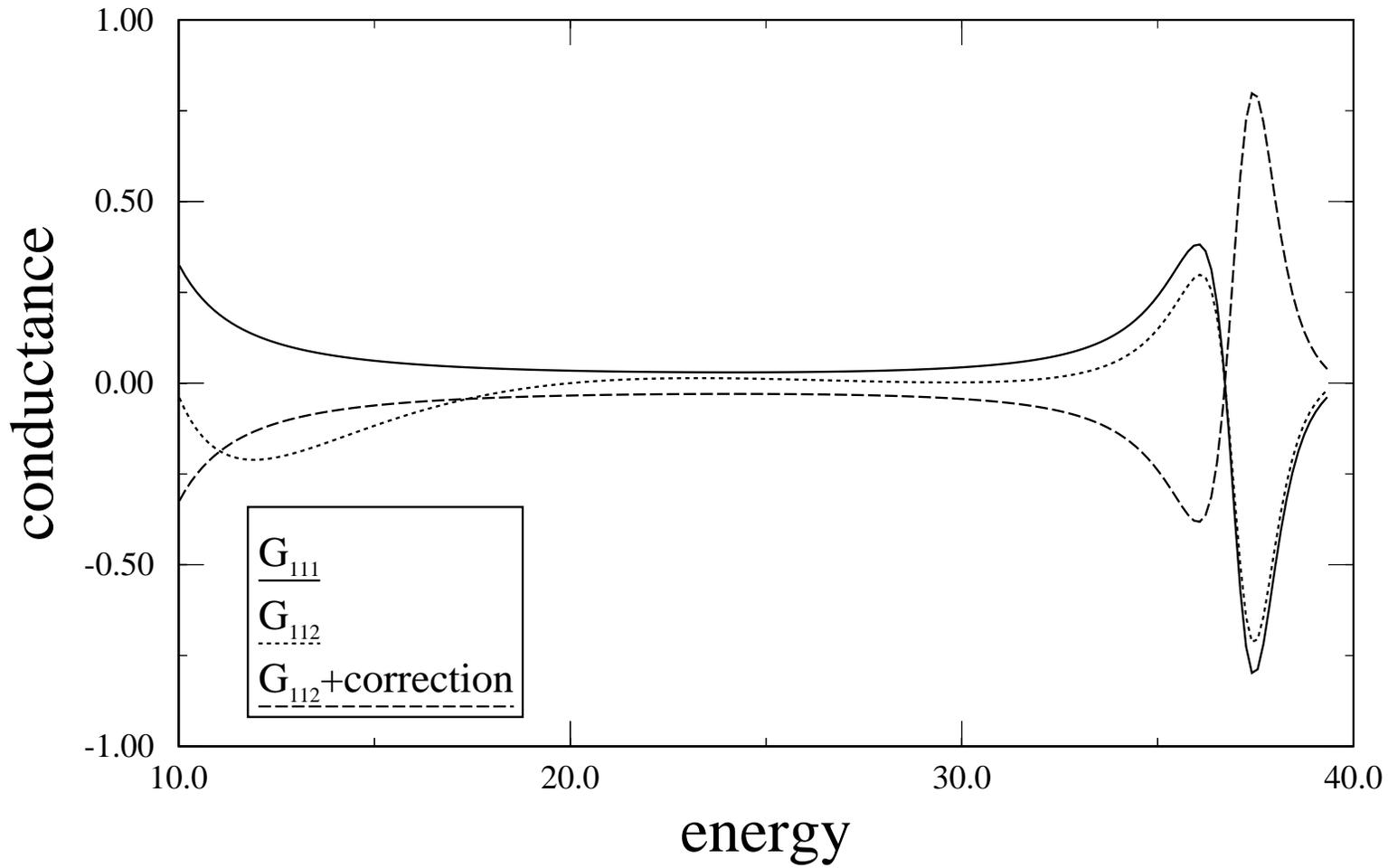
\end{document}